Volatility in the Relative Standard Deviation of Target Fulfilment as Key Performance Indicator (KPI)

Andreas Bauer[1], Jasna Omeragic[2]

[1] *Mendel University in Brno, Faculty of Business and Economics,*

*Provozně-ekonomická fakulta, Zemědělská 1, 613 00 Brno, Czech Republic*

[2] *Faculty of Electrical Engineering and Information Technologies,*

*SS Cyril and Methodius University in Skopje, North Macedonia*

*Corresponding author. E-Mail:* [a.bauer@netvance.com](a.bauer@netvance.com)

In this study, we identify the relative standard deviation volatility (RSD volatility) in the individual target time fulfilment of the complete set of comparables (e.g., all individuals in the same organisational structure) as a possible key performance indicator (KPI) for predicting employee job performance. KPIs are a well-established, measurable benchmark of an organisation's critical success metrics; thus, in this paper, we attempt to identify employees experiencing a transition in their RSD towards a higher per cent deviation, indicating emerging inadequate work conditions. We believe RSD volatility can be utilised as an additional assessment factor, particularly in profiling.

Introduction

Relative Standard Deviation (RSD) volatility in target achievements is a straightforward metric, and yet, it is critical for developing an optimal work environment. The target time of the indicator is a quantifiable feature, and key performance indicators (KPIs) are a measurable representation of an organisation's critical success factors. Thus, KPIs should be selected based on the context of the organisation. Each KPI should be aligned with and quantified against business objectives. KPIs aid in the definition of critical roles and the development of departmental performance indicators. Consequently, the performance of an organisation can be quantified. Performance management requires the establishment of specific and attainable KPIs.

*Benefits of this research*

The findings of this study can guide future research, in addition to creating a baseline for evaluating employees' working conditions where RSD volatility in target time fulfilment represents an intervening variable to consider. Company performance measures should be developed with a specific objective in mind and not chosen randomly, and this should naturally result in a minimum amount of performance measurements. Measuring the performance of a business in a disciplined manner would enable organisations to acquire genuine insights that will aid managers in making better decisions (Marr et al., 2004). As a result of this study, the institution might include fluctuation in target time fulfilment as an intervening variable for employees when developing a working environment and position guidelines.

*KPI in Balanced Scorecards (BSC)*

Based on our results, we recommend the practical use of RSD volatility, especially in balanced scorecards (Kaplan and Norton, 1996) in the segment of internal processes, instead of the position of employee fluctuation, which can only be included in the evaluation as a very late indicator. Our measurements show that the RSD volatility is already displayed beforehand in a significant impact. The KPI chosen for the BSC is the reduction in RSD volatility over the measurement period. We believe that the responsibility for the measurement usually lies with the operations manager role.

Kaplan and Norton designed the balanced scorecard to manage performance, offer management a summary of a company's KPI and simplify operations alignment with strategy. Their research impetus was that companies depended on standard financial accounting ratios, which gave a narrow and incomplete picture of the company's performance. The foundation of the balanced scorecard is integrating the efforts of the four areas in a causal chain that includes all four views: Kaplan and Norton emphasise that non-financial strategic objectives should not consist of an incredible collection of measures but should involve a balanced representation of financial and non-financial measures. Our proposed RSV volatility KPI has a firm grounding in the theoretical works of Kaplan and Norton (Kaplan et al.,1996) but provides a practical approach that the community can implement. We are considering examples of the BSC's practical application in business (Fernandes et al., 2006, Marr et al., 2004, Satria et al., 2021). The advantage of utilising a leading KPI (employee RSD volatility) rather than a lagging KPI (employee turnover) will become apparent.

*Profiling of employees*

Measuring intellectual capital is one of the companies most significant difficulties in the 21st century. A vital performance indicator should quantify an organisation's progress and help it define and evaluate success.

Especially when we consider the significance of quantifying organisational knowledge assets (Marr et al., 2004) show that knowledge assets underpin a company's capabilities and core competencies; therefore, they are the most important and must be measured.

As a component of workforce planning, profiling focuses on matching individuals to workloads and ensuring enough manpower to account for cyclical activity level volatility (Thain et al. 2002). Profiling methods are applied when quantifiable work volumes can be determined and anticipated correctly. We take RSD volatility as a supplemental assessment factor used in conjunction with other KPI factors, specifically in the context of profiling. A similar study that supported our method used logistic regression to analyse worker turnover (Setiawan et al., 2020).

Based on the specification of KPIs as a quantitative measure across time, Marr and Schiuma propose visualising the paths of knowledge assets to develop strategic KPIs that are easy to use and can be utilised to manage and report these crucial drivers in today's challenging economic situation (Marr et al., 2004).

*Daily fluctuations concerning monthly fluctuations*

In past studies, the focus was primarily on short-term daily influences. We see that, particularly in our RSD volatility model, overall impacts are uncovered. While overall work engagement levels are generally stable (e.g. Mauno et al. 2007), daily levels vary significantly (Sonnentag et al. 2012). At times, a highly motivated individual is disengaged from work, whereas an unmotivated individual under certain conditions can account for daily shifts in work commitment (Bledow et al. 2011). According to standard inventories, employees with low employee reliability scores were more hostile, impulsive, insensitive, self-absorbed, and unhappy in a 1989 historical study. (Hogan and Hogan 2018). However, these approaches are very much related to individual surveys and, therefore, cannot be used neutrally as a KPI in the same manner as the proposed RSD volatility KPI.

Research methodology

*Target hours VS achieved hours*

As a data source, we used daily performance assessment measurements from an internal enterprise resource planning (ERP) system: The productivity of individual employees (actual working times) and target working times (target times) were recorded. Our study utilised data on existing and target operating hours of 19 employees between January 2016 and October 2021.

*RSD anomalies in target time fulfilment*

We determine the relative standard deviation (RSD) we process $\sigma$, such that $RSD = \frac{\sigma}{\mu}$, where $\sigma = \sqrt{\sum\{x_{at} - \mu(x_{tt}; x_{at})^2\} + \{x_{tt} - \mu(x_{tt}; x_{at})^2\}}$, and $x_{tt}$ is the target time and $x_{at}$ is the actual time.

In our study, we define RSD volatility as a factor in which the response time value serves as a proxy for more in-depth analysis. A low RSD indicates that the data points are more stable or performance is consistent throughout the system. A large RSD, on the other hand, usually means that the data covers a wide range of values and that target time fulfilment is more unpredictable and unstable than total quantity. With the help of the RSD volatility method, we can identify the employees who are experiencing a transition in their RSD towards a higher percentage deviation, as shown in Figure 1 below. This is consistent with the findings of Nanda et al. (2020) on the effects of change in the unique work environment on performance. The study's limitation lies in the study's small sample size; the long observation period is an effort to compensate for this shortcoming.

Analysing the personal performance of each employee, we calculated the RSD separately. Using the RSD volatility as a factor, we can highlight the most successful employees presented on the chart whose result for RSD is about 10%.

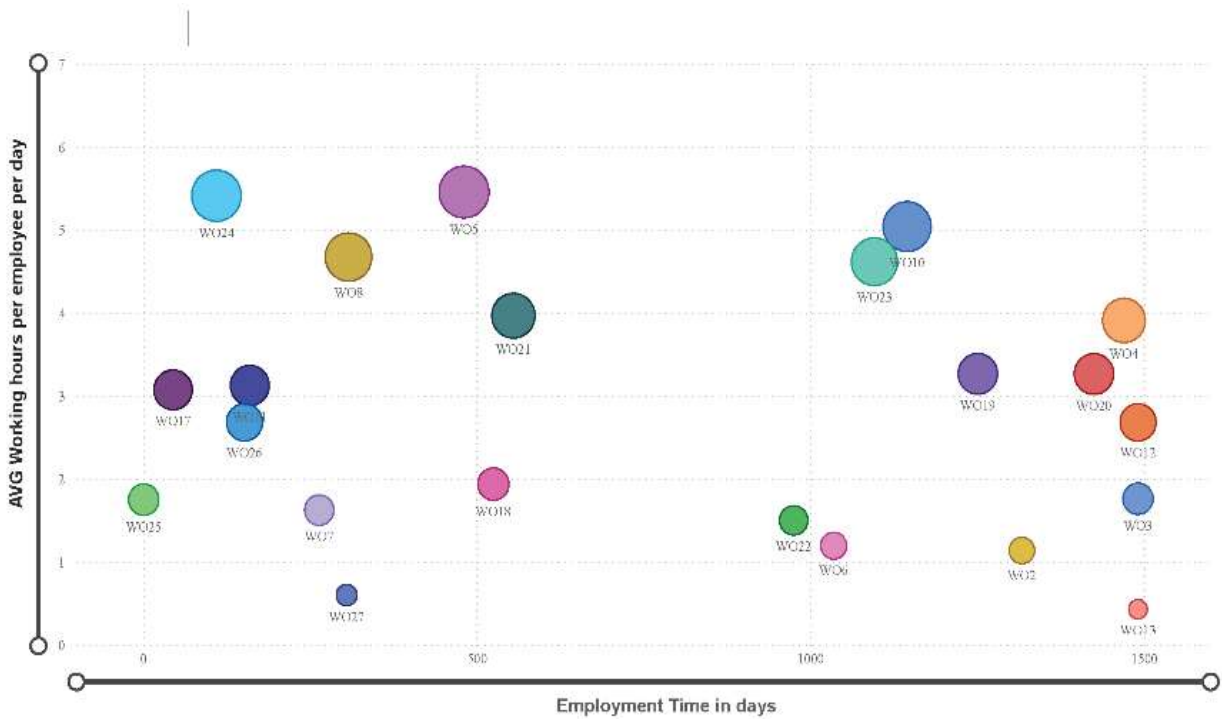

Figure 1. Sample data distribution: working time and duration of employment

The differences between $x_{at}$ and $x_{tt}$ are as follows: each employee $x_{tt}$ and $x_{at}$ are presented as the target time difference $\Delta x_{tt}$, and $\Delta x_{at}$ for the employee's first target time $x_{tt}^n$ and the previous $x_{tt}^{n-1}$. Using employee WO10 as a sample data point in Figure 2, employee WO10 has an output of 373.75h for $\Delta x_{at}$ and -416h for $\Delta x_{tt}$.

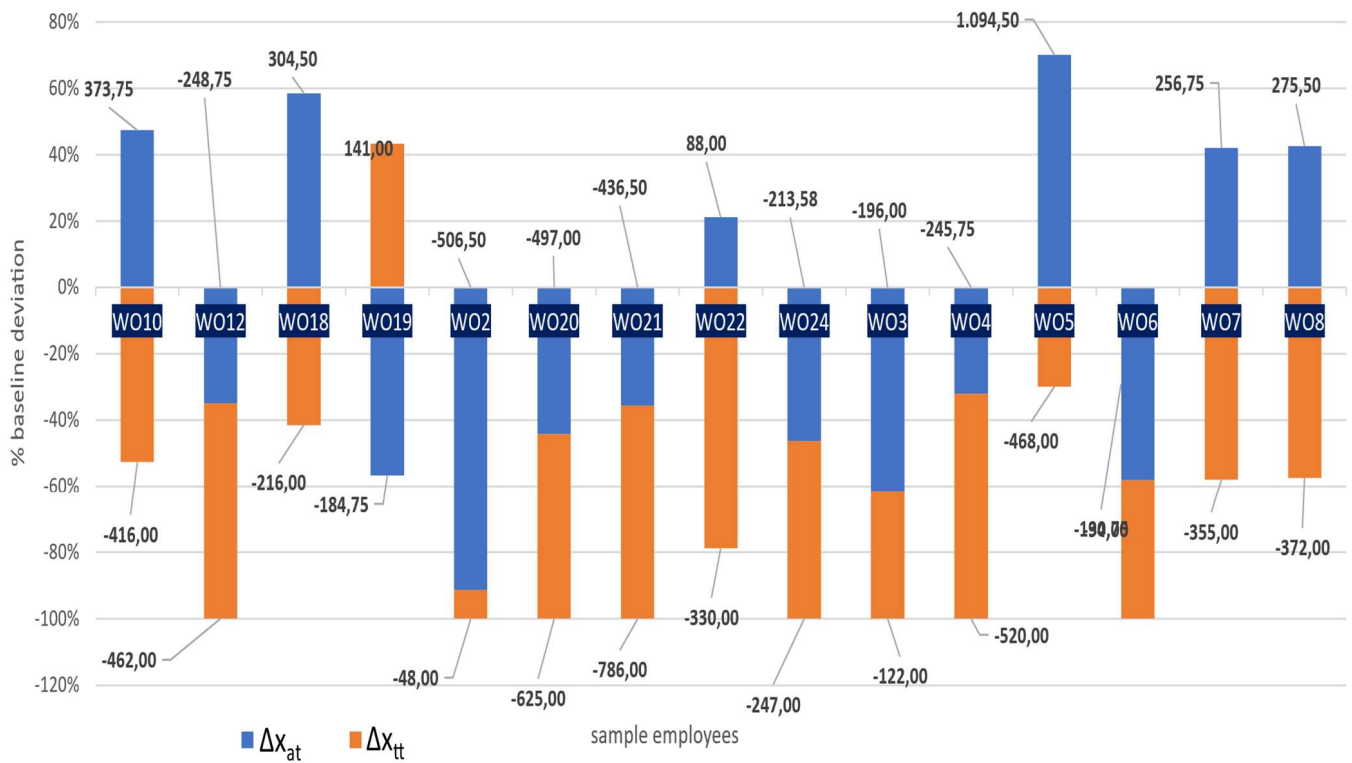

Figure 2. Data for the analysis of movement through years, showing spread in target time fulfilment for sample employees (WO01 to WO24)

The example of the evaluation of employee performance internal business process approach in the Sumatra Power Plant case study shows the advantage of using RSD Volatility. In the case study, the process is specified as follows: Internal Business Process Perspective aims to identify each composition performed and develop possible work measurement methods as control exposures that can improve and maintain the company's performance. By implementing our RSD Volatility, KPIs, up to four individual factors can be combined in an objectively measurable formula to form a KPI. They can thus be recorded in a predictable quality. This case study relates specifically to the following parameters:

Percentage of employee attendance rate (PBI1), number of employees who resigned (PBI2), Percentage of the Number of employees who fit the placement of their field (PBI3), number of late and often absent employees (PBI4) (Lubis et al., 2021).

Results

The proposed RSD volatility method identifies an indication of a decrease in an individual employee's hourly target achievement rate by visualising a shift in the RSD value to a more significant deviation value when using the hourly target achievement rate.

Using fluctuations in target time fulfilment as the RSD volatility method, we have established each employee's position (performance). We took into account fatigue, level of stress per day as well as the total work engagement time in the company and taking everything into account, we defined that the fulfilment of actual working hours and targets should be accompanied by a per cent of standard deviation, considered stable and good for the company. Then the standard deviation value was defined as a benchmark for the classification of stable and unstable job performance.

The fact that the total number of employees is considered in the equation means that parameters that affect the entire group are effectively filtered out, and the predictive accuracy of disturbing influences that only affect individual employees increases. Figure 3 is a graph for the evaluation of the sample data. One can see substantial volatility in the RSD of employees in the sample (WO03 to WO24) concerning the entire team over six years of observation.

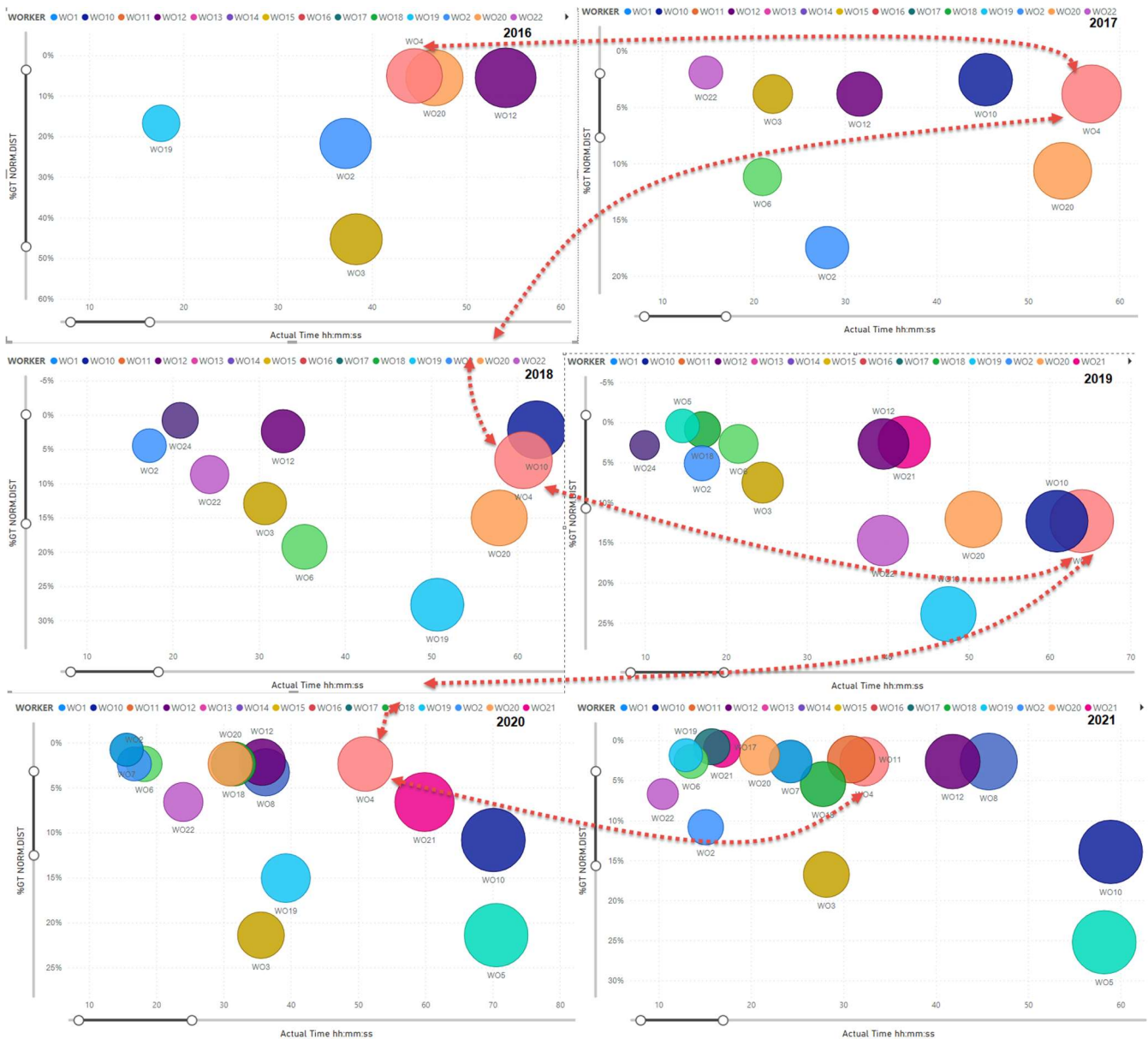

Figure 3. Data on RSD volatility of employees in the sample during an observation period of six years.

Conclusion

Our research demonstrates that a straightforward variable, such as the volatility of the target time fulfilment ratio, calculated as the relative standard deviation (RSD), can be incorporated into the key performance indicators (KPIs) used to assess employee performance predictability. On the one hand, the addition of target agreements with employees and the usage and application of KPIs to evaluate the efficiency and predictability of employees' work performance are now feasible using the RSD variation method.

Our proposed RSD volatility KPI can therefore be used to identify employee training potential and support person profiling. As an additional aid for employee analysis, it determines who possesses (and lacks) the abilities identified by the job–task analysis (Salas et al., 2012). With limitless resources, it may be beneficial to train everyone. Still, with restricted resources, the training can focus on those with the most significant gaps between their actual and required competencies by implementing the RSD volatility. Individual traits influencing the relative efficiency of various training programmes can also be examined using a more precise person analysis.

When the RSD variation increases, it can serve as an early warning system for potential vulnerabilities and emerging difficulties in an employee's work environment and function as an early trigger for implementing measures to improve operations.

References

Bledow, R., A. Schmitt, M. Frese, and J. Kühnel. 2011. "The Affective Shift Model of Work Engagement." *Journal of Applied Psychology* 96 (6): 1246–1257. https://doi.org/10.1037/a0024532.

Hogan, J., and R. Hogan. 2018. "How to Measure Employee Reliability." *Occupational Crime* 74 (2): 377–384. https://doi.org/10.4324/9781315193854-21.

Mauno, S., U. Kinnunen, and M. Ruokolainen. 2007. "Job Demands and Resources as Antecedents of Work Engagement: A Longitudinal Study." *Journal of Vocational Behavior* 70 (1): 149–171. https://doi.org/10.1016/j.jvb.2006.09.002.

Nanda, A., M. Soelton, S. Luiza, and E.T.P. Saratian. 2020. "The Effect of Psychological Work Environment and Work Loads on Turnover Interest, Work Stress as an Intervening Variable." Proceedings of the 4th International Conference on Management, Economics and Business (ICMEB 2019), 120, 225–231. https://doi.org/10.2991/aebmr.k.200205.040.

Setiawan, I., S. Suprihanto, A.C. Nugraha, and J. Hutahaean. 2020. "HR Analytics: Employee Attrition Analysis Using Logistic Regression." *IOP Conference Series: Materials Science and Engineering* 830 (3). https://doi.org/10.1088/1757-899X/830/3/032001.

Sonnentag, S., E.J. Mojza, E. Demerouti, and A.B. Bakker. 2012. "Reciprocal Relations Between Recovery and Work Engagement: The Moderating Role of Job Stressors." *Journal of Applied Psychology* 97 (4): 842–853. https://doi.org/10.1037/a0028292.

Marr, B., Schiuma, G., & Neely, A. (2004). Intellectual capital – defining key performance indicators for organizational knowledge assets. Business Process Management Journal, 10(5), 551–569. https://doi.org/10.1108/14637150410559225

Fernandes, K.J., Raja, V., Whalley, A.: Lessons from implementing the balanced scorecard in a small and medium size manufacturing organization. Technovation 26(2006), 623–634 (2006)

Kaplan, Robert S., and David Norton. "Using the Balanced Scorecard as a Strategic Management System." Harvard Business Review 74, no. 1 (January–February 1996): 75–85. (Reprint #96107.)

Salas, E., Tannenbaum, S. I., Kraiger, K., & Smith-Jentsch, K. A. (2012). The Science of Training and Development in Organizations: What Matters in Practice. Psychological Science in the Public Interest, Supplement, 13(2), 74–101. https://doi.org/10.1177/1529100612436661

Satria Lubis, A., Amalia, A., Ekonomi dan Bisnis, F., & Sumatera Utara, U. (2021). Employee Performance Assessment With Human Resources Scorecard And AHP Method (Case Study : PT PLN (PERSERO) North Sumatra Generation). Journal of Management Analytical and Solution, Vol. 1(2). https://talenta.usu.ac.id/jomas/article/view/6287